# Structural Resilience Analysis of an Internet Fragment Against Targeted and Random Attacks – A Case Study Based on iThena Project Data

Łukasz Świerczewski

Cyber-Complex Foundation, Poland

lswierczewski@cybercomplex.net

## Abstract

This article presents an analysis of the structural resilience of a fragment of Internet topology against both targeted and random attacks, based on empirical data obtained from the iThena project. Using a processed visualization of the network, a graph representing node connections was generated and subsequently subjected to detailed analysis using centrality metrics and community detection algorithms. Two attack scenarios were carried out: removal of nodes with the highest betweenness centrality and random removal of an equivalent number of nodes. The results indicate that targeted attacks have a significantly more destructive impact on the cohesion and functionality of the network than random disruptions. The article highlights the importance of identifying critical nodes and developing monitoring and protection mechanisms for Internet infrastructure in the context of cybersecurity.

## Streszczenie

W artykule przeprowadzono analizę odporności strukturalnej fragmentu topologii Internetu na ataki ukierunkowane i losowe, wykorzystując dane empiryczne pozyskane z projektu iThena. Na podstawie przetworzonej wizualizacji sieci wygenerowano graf reprezentujący połączenia między węzłami, który następnie poddano szczegółowej analizie z użyciem miar centralności oraz algorytmów wykrywania społeczności. Zrealizowano dwa scenariusze ataku: usunięcie węzłów o najwyższej centralności pośrednictwa oraz losowe usunięcie równoważnej liczby węzłów. Wyniki wskazują, że ataki ukierunkowane mają znacznie bardziej destrukcyjny wpływ na spójność i funkcjonalność sieci niż działania losowe. Artykuł podkreśla znaczenie identyfikacji krytycznych węzłów oraz rozwijania mechanizmów monitorowania i ochrony infrastruktury internetowej w kontekście bezpieczeństwa cybernetycznego.

**Słowa kluczowe:** BOINC, sieci komputerowe, Internet, grafy, cyberbezpieczeństwo, odporność

## Introduction

The modern Internet forms the backbone of global communication, connecting billions of devices and users worldwide. Its complex, distributed structure, while highly efficient, remains vulnerable to a range of structural threats, such as traffic exchange point failures, Denial of Service (DoS) attacks, or routing manipulations (e.g., BGP hijacking). Understanding the topology of the network and identifying its critical elements is essential not only for performance assurance but above all for maintaining the resilience of the information infrastructure.

In this study, we attempt to analyze the resilience of a fragment of Internet topology to both targeted and random attacks. The basis of the analysis is data sourced from the iThena project (ithena.net) – an experimental distributed computing platform that enables modeling and visualization of the real structure of the Internet through traceroute and ping measurements from various points in the network.

Based on a processed visualization of a network segment acquired from iThena, we constructed a graph representing the local structure of Internet connectivity. A detailed topological analysis of this graph was then performed, including node centrality, community detection, and global metrics such as modularity. The main objective was to compare the effects of removing nodes based on their topological importance (targeted attack) versus random node removal. The analysis aimed to determine which scenario poses a greater threat to the cohesion and functionality of the observed network.

The conducted experiments and obtained results provide arguments for the necessity of protecting the most critical elements of the Internet infrastructure, as well as for the development of mechanisms for its dynamic monitoring and incident response.

## Literature Review and Related Research

The structural analysis of computer networks and their resilience to disruption has been the subject of extensive scientific research over the past two decades. One of the key points of reference is the work of Barabási and Albert (1999), who described the scale-free network model, in which a small number of nodes have a very high number of connections. Such networks show significant resistance to random failures but are particularly vulnerable to targeted attacks on the most connected nodes.

Subsequent studies, such as those by Cohen et al. (2001), confirmed this dependency by presenting theoretical and simulation-based models of the impact of node removal on global network cohesion. The literature also highlights the importance of betweenness centrality as a metric for identifying critical points in a network's topology that play a vital role in data transmission (Freeman, 1977).

Studies based on real-world topological data—such as those conducted within the CAIDA, RIPE Atlas, or DIMES projects—have enabled empirical verification of theoretical assumptions about Internet structure. Particularly noteworthy are studies analyzing the effects of BGP (Border Gateway Protocol) manipulation, which indicate that even local routing changes can lead to serious service availability issues (Gill et al., 2006).

The iThena project aligns with this line of research by providing distributed real-time measurement data. Its use allows not only for the observation of dynamic changes in Internet topology but also for testing resilience under various disruption and attack scenarios. Previous research has mainly relied on simulated or centralized measurement data—iThena helps fill this gap by offering insights from a wide range of geographic locations.

In the context of this article, the presented approach aims to combine classical topological analysis with a practical assessment of structural network resilience based on empirical data, making it a valuable contribution to the advancement of research on Internet infrastructure security.

## Methodology

The study was based on an analysis of a fragment of the Internet topology graph obtained from the iThena project in the form of a network structure visualization. The topology data is from 2020. In the first stage, a static image representing the graphical structure of network connections was processed by identifying nodes (connection points) and estimating links between them using thresholding and contour analysis.

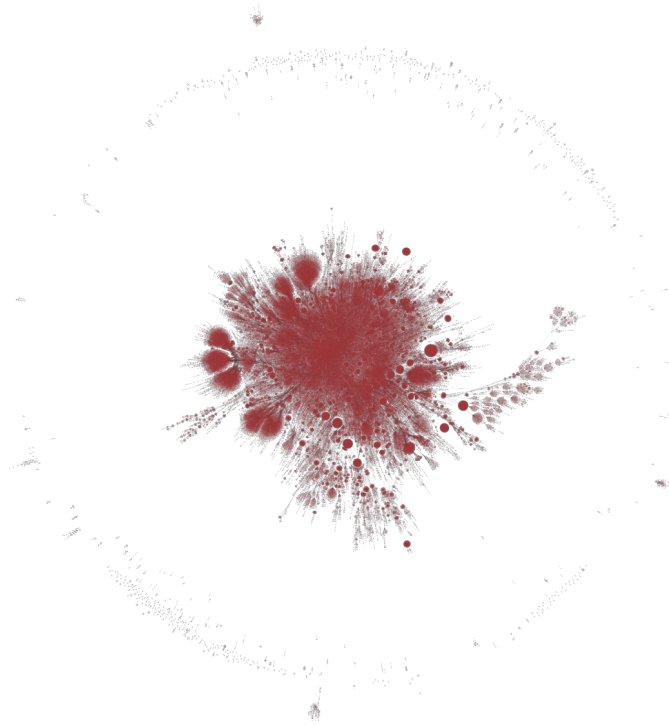

***Figure 1.*** *A large portion of the graph generated from 2020 data on the topology of the global Internet network*

Next, the identified nodes and edges were converted into graph format, allowing the use of classical network analysis methods. A central fragment with the highest density was extracted from the graph and used to perform a structural resilience analysis.

Three basic centrality measures were calculated for the nodes: degree centrality, betweenness centrality, and closeness centrality. Additionally, community detection algorithms were applied to identify the modular structure of the network. The modularity index was used to evaluate the network's cohesion.

Two attack scenarios were conducted as part of the experiments:

   a) **Targeted attack** – involved the removal of one-third of the nodes with the highest betweenness centrality;
   b) **Random attack** – involved randomly removing one-third of the nodes.

After each attack, the number of resulting connected components, the size of the largest component, and the average path length within the largest component were assessed. The comparison of these indicators enabled the evaluation of the impact of each attack type on the integrity and functionality of the network.

## Results

The analysis of the selected network topology fragment allowed for a clear distinction between the effects of a targeted and a random attack. In the case of a targeted attack on one-third of the nodes with the highest betweenness centrality, a significant drop in network cohesion was observed:

| Metric | Before Attack | After Targeted Attack | After Random Attack |
|---|---|---|---|
| Number of connected components | 29 | 67 | 40 |
| Size of the largest component | 441 | 123 | 291 |
| Average path length | 7.75 | 7.5 | 8.24 |

These results indicate that targeted attacks, even when removing the same number of nodes, have a significantly greater impact on the structure and functionality of the network. The fragmentation into many small components and the reduction of connectivity in the largest component considerably decrease the network's resilience and data transmission efficiency.

The analysis utilized several graph-based algorithms, including the Girvan–Newman algorithm for community detection, which identifies clusters by iteratively removing edges with the highest betweenness centrality. Centrality measures were computed using standard algorithms provided by the NetworkX library, enabling the identification of the most influential nodes. The attacks were simulated by systematically removing selected nodes and observing changes in the graph's structure.

Additionally, visualizations of the graphs before and after the attacks support these observations:

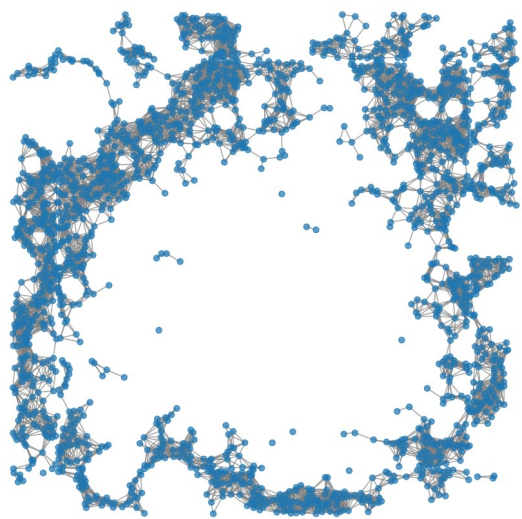

*Figure 2.* Graph before the attack

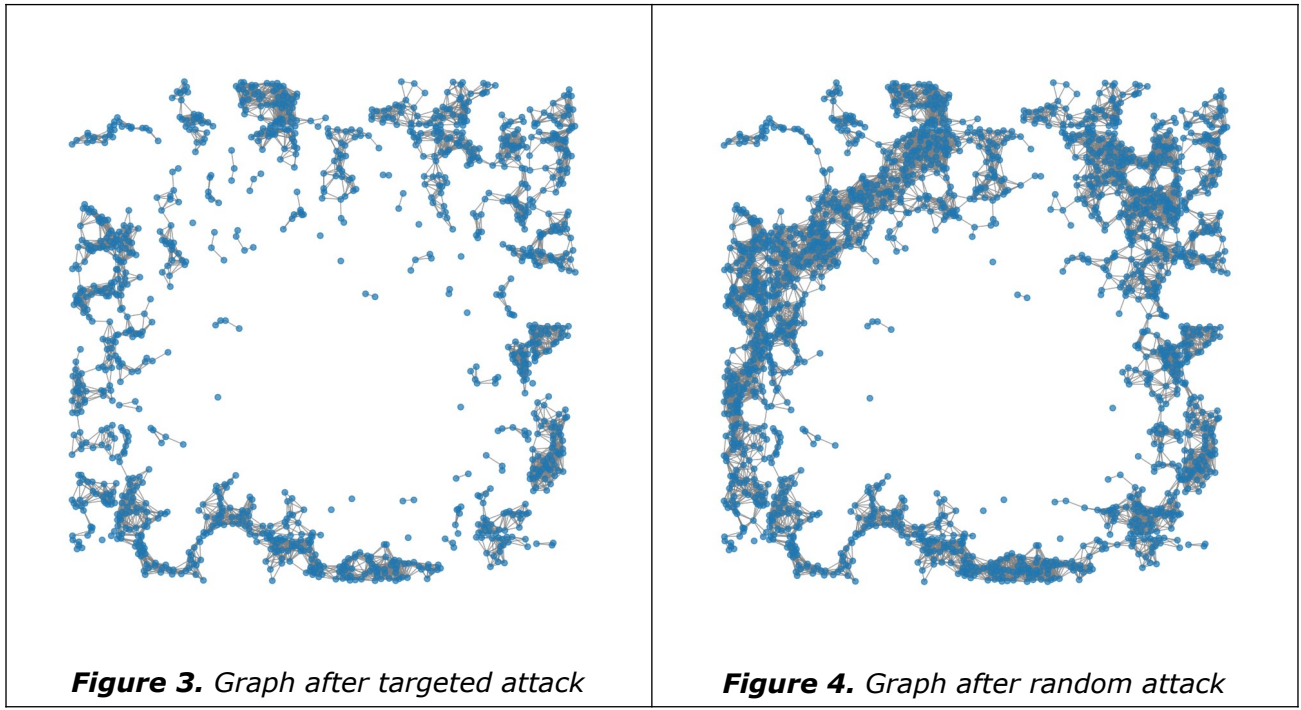

*Figure 3.* Graph after targeted attack       *Figure 4.* Graph after random attack

A strong fragmentation and dispersion of the structure is visible in the case of the targeted attack, with relative cohesion maintained following the random attack.

## Discussion

The obtained results have significant implications from the perspective of network infrastructure security. In particular, they demonstrate that even a partial attack directed at carefully selected nodes (e.g., those with high betweenness centrality) can lead to a considerable degradation in network cohesion. Such attacks can be carried out deliberately by actors possessing knowledge of the network structure—for instance, to disrupt communications, intercept traffic, or conduct disinformation operations.

The analysis shows that ensuring network resilience requires not only real-time monitoring but also proactive identification and protection of the most critical topological elements. Nodes acting as intermediaries in communication (with high betweenness centrality) should be prioritized for protection, both physically and logically.

At the same time, increasing structural redundancy is crucial, by building alternative paths and connections that enable traffic rerouting in the event of failure or attack. A well-designed topology can significantly enhance the network's resistance to disruptions, reducing the risk of complete fragmentation even under targeted sabotage.

In the context of an increasingly dynamic cybersecurity threat landscape, the results of this study indicate a strong need to further develop tools for simulation, analysis, and visualization of network topologies. Projects like iThena can serve as a valuable source of data for building early warning systems and automatically detecting structural vulnerabilities on a global scale.

## Conclusions

The conducted study demonstrated that the topological structure of a fragment of the Internet is particularly vulnerable to targeted attacks on key network nodes. The removal of just one-third of nodes with the highest betweenness centrality leads to a significant breakdown of the network and a decline in its ability to communicate efficiently. Compared to random attacks, the effects of targeted actions are significantly more destructive.

The obtained results confirm the necessity of developing methods for identifying and protecting critical points in the Internet topology. Furthermore, they highlight the usefulness of graph-based analysis as an effective tool supporting cybersecurity, network engineering, and critical infrastructure planning.

The use of data obtained from the iThena project shows that distributed and empirical measurement sources can serve as a solid foundation for the practical assessment of the structural resilience of networks. Further research is recommended, focusing on the analysis of other Internet segments, real-time attack scenario simulations, and the integration of such analyses with network security systems.

## References

1. Barabási, Albert-László, and Réka Albert. "Emergence of scaling in random networks." *science* 286.5439 (1999): 509-512.

2. Cohen, Reuven, et al. "Breakdown of the internet under intentional attack." *Physical review letters* 86.16 (2001): 3682.

3. Freeman, Linton C. "A set of measures of centrality based on betweenness." *Sociometry* (1977): 35-41.

4. Gill, Phillipa, Michael Schapira, and Sharon Goldberg. "A survey of interdomain routing policies." *ACM SIGCOMM Computer Communication Review* 44.1 (2013): 28-34.

5. Hagberg, A., & Conway, D. (2020). Networkx: Network analysis with python. *URL: https://networkx. github. Io*, 1-48.

6. Blondel, Vincent D., et al. "Fast unfolding of communities in large networks." *Journal of statistical mechanics: theory and experiment* 2008.10 (2008): P10008.

7. Newman, Mark EJ, and Michelle Girvan. "Finding and evaluating community structure in networks." *Physical review E* 69.2 (2004): 026113.

8. Anderson, David P. "Boinc: A system for public-resource computing and storage." *Fifth IEEE/ACM international workshop on grid computing*. IEEE, 2004.

9. Anderson, David P. "BOINC: a platform for volunteer computing." *Journal of Grid Computing* 18.1 (2020): 99-122.